# Wafer-scale CVD Growth of Monolayer Hexagonal Boron Nitride with Large Domain Size by Cu Foil Enclosure Approach

*Xiuju Song, Junfeng Gao, Teng Gao, Yufeng Nie, Jingyu Sun, Yubin Chen, Chuanhong Jin, Feng Ding*[*]*, Yanfeng Zhang*[*]*, Zhongfan Liu*[*]

X. J. Song, Dr. T. Gao, Y. F. Nie, Dr. J. Y. Sun, Y. B. Chen, Prof. Y. F. Zhang, Prof. Z. F. Liu
Center for Nanochemistry (CNC), Beijing Science and Engineering Center for Low Dimensional Carbon Materials, Beijing National Laboratory for Molecular Sciences, College of Chemistry and Molecular Engineering, Academy for Advanced Interdisciplinary Studies, Peking University, Beijing 100871, P. R. China
E-mail: zfliu@pku.edu.cn; yanfengzhang@pku.edu.cn

Dr. J. F. Gao, Prof. F. Ding
Institute of Textiles and Clothing, Hong Kong Polytechnic University, Hong Kong, P. R. China
E-mail: feng.ding@polyu.edu.hk

Prof. C. H. Jin
State Key Laboratory of Silicon Materials, Key Laboratory of Advanced Materials and Applications for Batteries of Zhejiang Province, Department of Materials Science and Engineering, Zhejiang University, Hangzhou 310027, P. R. China

Keywords: hexagonal boron nitride, chemical vapor deposition, Cu foil, large domain size, orientation

Two dimensional materials have received increasing attentions since the discovery of graphene.[1] Specifically, hexagonal boron nitride (*h*-BN), a structural analogue of graphene, possesses only 1.8% lattice mismatch with graphene but a large band gap (~5.9 eV). The combinations of graphene and *h*-BN, including both in-plane *h*-BN-graphene hybrids and stacked graphene/*h*-BN (G/*h*-BN) structures, have been demonstrated very intriguing physical properties such as Hofstadter's butterfly.[2-6] In particular, G/*h*-BN vertical stacks exhibit the best field effect transistor performance with extremely high carrier mobility.[7] As the perfect dielectric layer, *h*-BN has an atomically flat and dangling bond-free surface, which ensures no charge traps at the G/*h*-BN interface and results in an order of magnitude increase of graphene's carrier mobility as compared to the typical $SiO_2$/Si substrate.[8, 9] *h*-BN has also stimulated various applications in deep ultraviolet light emitters,[10] protective coatings,[11] and transparent electronics,[12] due to its excellent mechanical strength, chemical inertness as well as nice optical transparency. These attractions of *h*-BN ignited the efforts on its synthesis in the last few years, targeting uniform thickness, large domain size and high crystallinity.

Indeed, the chemical vapor deposition (CVD) growth of *h*-BN has been explored by using a variety of transition metals as substrates.[13-16] Cu foil is the typical substrate for *h*-BN growth due to its low cost, commercially easy availability and well-behaved catalytic performance for high-quality *h*-BN film. Kim *et al.* pioneered the monolayer *h*-BN growth on Cu foil *via* a low pressure CVD (LPCVD) route and obtained maximum *h*-BN triangles with ~1 μm in edge length.[17] By employing electropolished Cu foils, Teo *et al.* obtained *h*-BN hexagons with a maximum edge length of ~5 μm very recently, attributable to the reduced nucleation density.[18] The nucleation sites can also be suppressed simply by increasing the pre-annealing time of Cu foil up to 6 hours, which gave the largest *h*-BN monolayer triangles with an edge length of ~20 μm.[19] The challenging issues remained on *h*-BN/Cu synthesis include the uniformity control, thickness control and larger domain size, which are crucial for various applications of the *h*-BN films, especially in high-performance G/*h*-BN devices.

In this work, we demonstrate the largest single crystalline domain *h*-BN monolayer on Cu foils by using LPCVD technique. The key of the success is to effectively suppress the nucleation density during CVD growth process by using the inner surface of folded Cu foil enclosure. Triangular-shaped single crystal *h*-BN flakes with a domain size up to 72 μm in edge length and a high monolayer percentage up to 92% have been obtained through the kinetic control of Cu-CVD process. More importantly, we found, for the first time, that the orientations of the as-grown *h*-BN flakes are strongly correlated to the underlying Cu crystalline facets. In other words, the symmetry of the Cu facet (representatively (111), (110), or (100)) greatly affects the orientations of *h*-BN monolayer on it. The Cu (111) facet is recognized as the ideal surface for the synthesis of well-aligned *h*-BN domains, which has two equivalent orientations only. The facet dependence growth behavior is well understood by the density functional theory (DFT) calculations that the edge of an *h*-BN domain tends to be well aligned along a high symmetric direction of the catalyst surface. The present study certainly provides the future directions for growing high-quality *h*-BN films by deepening the fundamental understanding of Cu-CVD process, which further opens a practical pathway for high-performance G/*h*-BN electronics.

As schematically illustrated in **Figure 1a**, *h*-BN was grown on polycrystalline Cu foils *via* a LPCVD method by using ammonia borane ($BH_3$-$NH_3$) as the precursor. Prior to the CVD growth, the Cu foil was electropolished to reduce the surface roughness and remove any attached contaminations, followed by folding it into an enclosure shape (**Figure S1**). This enclosure ensured a far cleaner growth environment on its inner surface and effectively suppressed the formation of BN nanoparticles frequently seen on its outer surface (**Figure S2**). The $BH_3$-$NH_3$ precursor was put into a specially-designed half-opened quartz cell and then loaded into the CVD growth tube, where a heating belt was wrapped around to aid the sublimation of $BH_3$-$NH_3$ at a temperature range of 65 °C ~ 120 °C. The $BH_3$-$NH_3$ was sublimated and decomposed into $(BH_2NH_2)_n$, $(BHNH)_3$ and hydrogen ($H_2$),[20] which were further pyrolysed into B- and N-containing intermediate species at the hot zone of growth tube for h-BN growth. Before feeding with the precursors, the Cu enclosure was annealed at 1000 °C for 2 hours in a flow of 20 sccm $H_2$ and 50 sccm Ar. Monolayer h-BN flakes with triangle shapes are obtained on the inner surface of Cu enclosure *via* a surface catalysis process (**Figure 1b**).

SEM micrograph in **Figure 1c** depicts the h-BN triangular flakes grown for 120 min at 1000 °C with a precursor heating temperature of 65 °C, which show a darker contrast with regard to the bare Cu substrate. The middle triangle presents a maximum edge length of ~72 μm, the largest single crystalline domain compared with those reported in recently published work.[19] These individual flakes gradually merge into a continuous layer with the increase of growth time, enabling the formation of full coverage of h-BN monolayer film on Cu foils. The optical microscope (OM) image in **Figure 1d** displays an h-BN film after transferred onto 280 nm $SiO_2$/Si *via* a conventional wet etching technique.[21] A uniform contrast can be seen for the h-BN covered area, suggesting the formation of uniform h-BN layer. The atomic force microscopy (AFM) images of the transferred sample reveal a thickness of ~ 0.7 nm, as shown in **Figure 1e**. The surface roughness of h-BN film is measured to be 0.145 nm, much lower than that of $SiO_2$/Si substrate (0.166 nm).[17, 22] Further statistical analysis of AFM height distribution (48 points in total, **Figure S3**) manifests a thickness fluctuation between 0.5~0.9 nm, well consistent with that of the typical monolayer h-BN.[18] These observations confirmed the monolayer nature of the CVD-grown *h*-BN film. Wafer-scale *h*-BN film has been grown in such a way with over 92% monolayer (**Figure 1f and Figure S3**).

Spectroscopic characterizations, including X-ray photoemission spectroscopy (XPS), Raman spectroscopy and UV-visible spectroscopy (UV-Vis), were carried out for obtaining the element composition and stoichiometry, lattice vibration modes as well as the band gap information of the obtained monolayer *h*-BN. Two characteristic XPS peaks located at 398.1 eV and 190.5 eV, which can be assigned to N 1s and B 1s signals, respectively, were observed. The N/B ratio is estimated to be 1.09, indicative of the predominant B-N chemical bonding in the *h*-BN lattice (**Figure 1g and Figure S4**). **Figure 1h** shows the Raman spectrum of *h*-BN film on SiO$_2$/Si substrate. The characteristic peak located at ~1370 cm$^{-1}$ is typical of monolayer *h*-BN, originating from the boron-nitrogen bond stretching.[23] The full width at half maximum (FWHM) of the peak is ~25 cm$^{-1}$, also consistent with the monolayer feature and the high crystallinity.[31] The optical band gap measured from UV-Vis spectroscopy of the *h*-BN monolayer transferred onto quartz substrate is 5.9 eV as shown in **Figure 1i**, which is very close to the theoretical value (6.0 eV).[24]

It is a general trend that, the *h*-BN film grown on Cu foils is of polycrystalline nature with small grains and highly concentrated grain boundaries and defects.[19] This is attributed to the extremely large nucleation centers created in the CVD growth process. Compared with the Cu-CVD-graphene process, the nucleation density on Cu foils during *h*-BN growth is generally much higher, which is most possibly attributed to the high chemical affinity of N-containing intermediate species to the Cu surface.[25, 26] Many factors determine the nucleation density on Cu foil surface, including precursor evaporation temperature and rate, surface roughness, growth temperature, and external impurities. Our systematic investigations indicate that the feeding rate of precursors is a key factor for nucleation density control. As solid precursor source is used in the present experimental setup, the feeding rate can be adjusted by varying the heating temperature of the precursor cell. **Figure 2a-c** exhibit the SEM images of *h*-BN monolayer triangles obtained on the inner surface of Cu enclosure at different precursor evaporation temperatures and hence feeding rates. At a precursor heating temperature of 120 $^o$C, 2 min growth resulted in most-available *h*-BN triangles with an average edge length of 2.8 μm, corresponding a nucleation density of ca. $1.7 \times 10^{-1}$ μm$^{-2}$ (**Figure 2a**). Simply increasing the growth time merely led to continuous films and even multilayers. When reducing the precursor heating temperature to 70 $^o$C, the size of the *h*-BN triangles increased to ~9.2 μm in edge length (**Figure 2b**) with a growth time of 90 min. Further reducing the precursor heating temperature to 55 $^o$C, almost no *h*-BN was observed within a growth time up to 3h (**Figure S5**). The optimized precursor evaporation temperature was 65 $^o$C, with which large *h*-BN triangles of ~50 μm in edge length are synthesized in 120 min (**Figure 2c**). The nucleation density in this case was estimated to be $2.7 \times 10^{-3}$ μm$^{-2}$, about two orders of magnitude lower than that with 120 $^o$C as the precursor evaporation temperature. Under the optimized growth condition, the coalescence of *h*-BN flakes on Cu foils could be realized by extending the growth time, enabling the formation of continuous monolayer film with a full surface coverage as seen in **Figure 2d.**

The above experimental observations strongly suggest that controlling the precursor feeding rate is crucial for suppressing the nucleation of *h*-BN. Indeed, the use of Cu enclosure approach is the key for drastically decreasing the feeding rate of precursor species by a few orders of magnitude down. As a comparative study, we examined the growth of *h*-BN on the outer surface of Cu enclosure. As shown in **Figure 2e** (see also **Figure S2**), numerous BN nanoparticles and small irregularly-shaped *h*-BN flakes were observed. In contrast, the inner surface of Cu enclosure gives very clean large *h*-BN triangular monolayers flakes at the same time (**Figure 2f** and **Figure S2**). Although there is a small window to reduce the precursor feeding rate by decreasing its evaporation temperature and optimizing other experimental parameters, very limited improvements on growth quality have been achieved using the conventional Cu foil method. A frequently-observed phenomenon is that, at relatively low feeding rate of precursors, almost no *h*-BN flakes could be seen on the outer surface of Cu enclosure in spite of the large *h*-BN formation on the inner surface (**Figure S2**). This is believed to arise from the difference of hydrogen etching effect. In fact, the *h*-BN formation reaction is always accompanied with the reverse hydrogen etching reaction.[27] Apparently, the hydrogen etching reaction on the outer surface exposed to the Ar/$H_2$ carrier gas is much faster than that on the inner surface. As a result, relatively high precursor feeding rate is necessary for *h*-BN growth to surpass the hydrogen etching reaction, which leads to the formation of small *h*-BN monolayer and few layer flakes. In brief, the Cu enclosure approach has the following three advantages for achieving high-quality *h*-BN monolayer growth: (a) drastically reducing the feeding rate of precursors and hence the nucleation density; (b) effectively suppressing the hydrogen etching reaction; and (c) preventing the Cu surface from external contaminations.

Transmission electron microscopy (TEM) combined with selected area electron diffraction (SAED) and electron energy loss spectroscopy (EELS) were employed to probe the layer thickness, crystallinity, and elemental stoichiometry of the obtained *h*-BN flakes. **Figure 3a** displays a bright-field TEM image of an *h*-BN triangle (with edge length of ~40 μm) transferred onto TEM grid. SAED patterns recorded at five random positions (marked by red dots in **Figure 3a**) on this *h*-BN triangle were overlaid with an image processing tool into one frame shown in **Figure 3b.** Apparently, the only one set of six-fold symmetric diffraction pattern with sharp spots justifies the large area uniform single crystalline nature.[19] Moreover, the high-resolution TEM image on the film edge (**Figure 3c**) with a line shape contrast also supports the single layer feature of the as-grown *h*-BN film. To investigate the chemical composition of the sample, EELS spectrum was recorded (**Figure 3d**) and the representative peaks of boron and nitrogen K-shell ionization edges with the characteristic $\pi^*$ and $\sigma^*$ energy loss peaks at boron and nitrogen are shown, indicating that the *h*-BN flake consists of $sp^2$ hybridization bonds.[28]

Dark-field TEM (DF-TEM) was then employed to examine the uniformity and orientation of *h*-BN triangular flakes. **Figure 3e** displays a false color DF-TEM image of an equilateral *h*-BN triangle with all perfect interior angles of 60°, indicative of its single crystalline nature. During CVD growth, such kinds of *h*-BN triangles gradually expand their sizes and finally merge with each other, forming an entire film. Apparently, the orientations of these triangles will determine the merging boundaries and the finally-formed polycrystalline films. One of the typical merging processes is shown in **Figure 3f**, in which two perfectly aligned flakes coalesce together. The corresponding overlaid SAED pattern exhibits only one set of hexagonal spots (**Figure 3g**), suggesting the identical crystalline orientation of these two domains. Another frequently-observed case for merging is the coalescence of two triangles with 180° of relative rotation to form a mirror-twinned structure (**Figure 3h**).[29] Their SAED patterns surprisingly showed only one set of hexagonal spots as seen in **Figure 3i**, indicating that the two triangles with an edge angle of 60° can also align precisely with each other. The above experimental observations suggest that the *h*-BN triangles with the same orientation or with a relative rotation of 180° are able to coalesce into a well-aligned film, which contribute to the high crystal quality with reduced grain boundaries. On the other hand, misoriented polygonal *h*-BN flakes can also be occasionally detected as shown in **Figure 3j-m**. The DF-TEM images clearly displayed two single crystalline domains with different orientations formed after merging (marked in different colors). The relative rotation of two merged triangles can be determined from the corresponding SAED pattern, which are ~21° in **Figure 3k** and ~14° in **Figure 3m**, respectively. Apparently, grain boundaries are created between these misaligned domains, leading to the polycrystalline nature of the *h*-BN film. It is hence a natural conclusion that a highly-aligned single crystalline *h*-BN monolayer film can be achieved by controlling the orientations of individual triangles in some way.

The orientation of adlayer is usually strongly affected by the crystalline facet of substrate.[30] The dependence of *h*-BN growth behavior on different crystalline facets of Cu foil was systematically examined by employing EBSD and SEM. **Figure 4a** shows the EBSD map of a typical area on Cu foil (corresponding SEM image shown in **Figure 4b**), which displays the coexistence of Cu (111), Cu (100) and Cu (110) facets, consistent with the XRD result shown in **Figure 4c**. The growth behaviors of *h*-BN films on these crystalline facets are exhibited in **Figure 4d-f**, respectively. Obviously, the nucleation densities and domain sizes of *h*-BN islands did not show remarkable facets dependent behavior. However, the orientation distributions of *h*-BN triangles are distinctly different (**Figure 4g-i**). The *h*-BN triangles are well aligned on Cu (111), where the difference of relative orientation can be expressed as multiples of $60°$ (**Figure 4d**). The two dominant orientations of *h*-BN on Cu (111) are $0°$ and $60°$, as clearly seen from the orientation distribution plot (**Figure 4g**). This indicates that the *h*-BN flakes are well aligned along the lattice of the Cu (111) surface. Contrastively, on Cu (100), there are four typical orientations, i.e. $0°$, $30°$, $60°$, $90°$ (**Figure 4e and h**) with the alignment deviations of $±30°$. Similarly, the *h*-BN flakes on Cu (110) facet have six dominant orientations with the deviations of $±10°$ (**Figure 4f and i**). These facet dependent orientations of *h*-BN on the Cu (100) and Cu (111) facets can be attributed to the alignment of the *h*-BN on the 4-fold and 6-fold symmetries of facets. However, the reason for the six orientations on Cu (110) is not clear, possibly caused by the reconstruction of Cu (110) facet under the experimental condition.

To achieve a deeper understanding of how the symmetry of the Cu facet affects the orientation of the grown *h*-BN, the interaction between the *h*-BN flakes and Cu facets was studied by DFT calculations (See Supporting information for details of the computation). It is known that the inert *h*-BN wall interacts weakly with the substrate surface (e.g., the calculated height of *h*-BN monolayer to Cu (111) surface is about 3.17 Å and the *van der Waals* interaction between them is only 0.104 eV per atom according to our calculations) and thus its orientation should be determined by the edge-catalyst interaction during the early stage of its growth, similar as that for graphene CVD growth.[31] The binding energy of an *h*-BN triangle with the substrate is the summation of the binding energies of its three edges, i.e. the perimeter of the *h*-BN triangle,

$$E(\theta) = Ee(\theta) + E(\theta + 2\pi/3) + E(\theta + 4\pi/4) \quad (1)$$

where E(θ) is the total bind energy of an *h*-BN flake on substrate; $Ee(\theta)$, $E(\theta + 2\pi/3)$, $E(\theta + 4\pi/4)$ are the binding energies of its three edges as shown in **Figure 5c**. θ, θ + 2π/3, θ + 4π/3 are the angles between the three edges (A, B, C) and one high symmetric direction of Cu surface, respectively. As shown in **Figure 5 i** and **j**, θ is the angle between edge A and the [$\bar{1}$ 1 0] direction of Cu (111) or [0 1 $\bar{1}$] direction of Cu (100).

Using DFT calculations, the binding energies of several *h*-BN edges with different angles i.e., θ = 0.0°, 8.1°, 18.4°, 26.6°, 31.0°, 36.9°, 39.8°, 45.0° on Cu (100) and θ = 0.0°, 6.6°, 19.1°, 23.4°, 26.3°, 30.0° on Cu (111), were calculated and shown in **Figure S6-S7**. The binding energies of the *h*-BN edge with other angles are obtained by linear interpolation method and those with angles beyond this range (0°-45° for Cu (100) surface and 0°-30° for Cu (111) surface) can be deduced by considering the symmetry of the system. Then the binding energy of an *h*-BN flake on Cu surface is calculated by Equation 1. As aforementioned, the equilateral triangular *h*-BN flake has an period of 120°, $E(\theta) = E(\theta + i \times 120°)$. Thus, as shown in **Figure 5a and b**, the binding energies of *h*-BN flakes on Cu (111) and Cu (100) to its perimeter only in the range of 0°-120° are plotted. It is distinct that there are two binding energy peaks for an *h*-BN triangle on the Cu (111) (0° and 60°) and four peaks on the Cu (100) ((0°, 30°, 60° and 60°), in good agreement with the experimental observations. The orientations of *h*-BN flakes corresponding to these peaks are schematically shown on the right panel of **Figure 5**. It is apparent that all these *h*-BN flakes possess at least one edge parallel to the close-packing direction of Cu surface (labeled with white lines). This can be understood that the close-packing direction has much dense Cu atoms, which passivate the edge of *h*-BN more effectively. Therefore, the highly populated orientation angles of *h*-BN flakes on Cu (111) and Cu (100) surfaces can be interpreted as the energetically preferred orientation of the *h*-BN flakes. This indicates that the growth of *h*-BN on Cu foil follows the edge-epitaxial growth mode,[32] which may be benefited from the low pressure environment on the clean inner cavity of Cu enclosure.

Such a conclusion allows us to achieve the *h*-BN orientation control by catalyst surface engineering. Such as on the Cu (111) surface, the *h*-BN layers has only two types of domains and the grain boundaries between them are mostly with a 60 degree mismatching angle or with a linear chain of B-B or N-N pairs.[33] And if one of the orientations could be suppressed greatly, e.g., by applying external field, breaking the symmetry of the surface *via* formation of metal steps, etc., then nearly all the *h*-BN flakes could be equally orientated on the Cu (111) surface and *h*-BN film formed by the coalescence of these flakes on a Cu (111) surface would be single crystalline.

In summary, high-quality *h*-BN monolayer films with the largest single crystalline domain size up to ~72 μm in edge length have been achieved on Cu foils using LPCVD technique. The folded Cu enclosure approach has been proved to perfectly suppress the nucleation centers and the reverse hydrogen etching reaction during CVD growth process, which leads to the remarkable improvement of single crystalline domain size and preferential monolayer growth. It is revealed, for the first time, that the orientations of as-grown *h*-BN monolayers are strongly correlated with the crystalline facets of growth substrates, with the Cu (111) being the best substrate for growing high-quality single crystalline *h*-BN monolayer films. DFT calculations well explained these crystalline facet effects. The present work provided the future direction for growing high-quality *h*-BN monolayer films and opens a practical pathway for high-performance G/*h*-BN electronics.

**Experimental Section**

The growth of *h*-BN film was performed on copper foils by using a low pressure chemical vapor deposition technique. (see also the Materials and Methods section in the Supporting Information)

**Supporting Information**

Supporting Information is available from the Wiley Online Library or from the author.

**Acknowledgements.** The work was supported by the Natural Science Foundation of China (Grants 51432002, 50121091, 51290272, 51222201), the Ministry of Science and Technology of China (Grants 2013CB932603, 2012CB933404, 2011CB933003, 2011CB921903, 2012CB921404), and the Ministry of Education (Grant 20120001130010).

Received: ((will be filled in by the editorial staff))
Revised: ((will be filled in by the editorial staff))
Published online: ((will be filled in by the editorial staff))

**References**

[1] K. S. Novoselov, A. K. Geim, S. V. Morozov, D. Jiang, Y. Zhang, S. V. Dubonos, I. V. Grigorieva, A. A. Firsov, *Science* **2004**, *306*, 666.

[2] B. Hunt, J. D. Sanchez-Yamagishi, A. F. Young, M. Yankowitz, B. J. LeRoy, K. Watanabe, T. Taniguchi, P. Moon, M. Koshino, P. Jarillo-Herrero, R. C. Ashoori, *Science* **2013**, *340*, 1427.

[3] W. Yang, G. Chen, Z. Shi, C.-C. Liu, L. Zhang, G. Xie, M. Cheng, D. Wang, R. Yang, D. Shi, K. Watanabe, T. Taniguchi, Y. Yao, Y. Zhang, G. Zhang, *Nat. Mater.* **2013**, *12*, 792.

[4] M. P. Levendorf, C. J. Kim, L. Brown, P. Y. Huang, R. W. Havener, D. A. Muller, J. Park, *Nature* **2012**, *488*, 627.

[5] Z. Liu, L. Ma, G. Shi, W. Zhou, Y. Gong, S. Lei, X. Yang, J. Zhang, J. Yu, K. P. Hackenberg, A. Babakhani, J.-C. Idrobo, R. Vajtai, J. Lou, P. M. Ajayan, *Nat. Nanotech.* **2013**, *8*, 119.

[6] M. Wang, S. K. Jang, W. J. Jang, M. Kim, S. Y. Park, S. W. Kim, S. J. Kahng, J. Y. Choi, R. S. Ruoff, Y. J. Song, S. Lee, *Adv. Mater.* **2013**, *25*, 2746.

[7] C. R. Dean, A. F. Young, MericI, LeeC, WangL, SorgenfreiS, WatanabeK, TaniguchiT, KimP, K. L. Shepard, HoneJ, *Nat. Nanotech.* **2010**, *5*, 722.

[8] R. Decker, Y. Wang, V. W. Brar, W. Regan, H. Z. Tsai, Q. Wu, W. Gannett, A. Zettl, M. F. Crommie, *Nano Lett.* **2011**, *11*, 2291.

[9] J. Xue, J. Sanchez-Yamagishi, D. Bulmash, P. Jacquod, A. Deshpande, K. Watanabe, T. Taniguchi, P. Jarillo-Herrero, B. J. LeRoy, *Nat. Mater.* **2011**, *10*, 282.

[10] K. Watanabe, T. Taniguchi, H. Kanda, *Nat. Mater.* **2004**, *3*, 404.

[11] Z. Liu, Y. Gong, W. Zhou, L. Ma, J. Yu, J. C. Idrobo, J. Jung, A. H. MacDonald, R. Vajtai, J. Lou, P. M. Ajayan, *Nat. Commun.* **2013**, *4*, 2541.

[12] G.-H. Lee, Y.-J. Yu, X. Cui, N. Petrone, C.-H. Lee, M. S. Choi, D.-Y. Lee, C. Lee, W. J. Yoo, K. Watanabe, T. Taniguchi, C. Nuckolls, P. Kim, J. Hone, *ACS Nano* **2013**, *7*, 7931.

[13] L. Song, L. Ci, H. Lu, P. B. Sorokin, C. Jin, J. Ni, A. G. Kvashnin, D. G. Kvashnin, J. Lou, B. I. Yakobson, P. M. Ajayan, *Nano Lett.* **2010**, *10*, 3209.

[14] Y. Shi, C. Hamsen, X. Jia, K. K. Kim, A. Reina, M. Hofmann, A. L. Hsu, K. Zhang, H. Li, Z. Y. Juang, M. S. Dresselhaus, L. J. Li, J. Kong, *Nano Lett.* **2010**, *10*, 4134.

[15] C. Zhang, L. Fu, S. Zhao, Y. Zhou, H. Peng, Z. Liu, *Adv. Mater.* **2014**, *26*, 1776.

[16] Y. Gao, W. Ren, T. Ma, Z. Liu, Y. Zhang, W.-B. Liu, L.-P. Ma, X. Ma, H.-M. Cheng, *ACS Nano* **2013**, *7*, 5199.

[17] K. K. Kim, A. Hsu, X. Jia, S. M. Kim, Y. Shi, M. Hofmann, D. Nezich, J. F. Rodriguez-Nieva, M. Dresselhaus, T. Palacios, J. Kong, *Nano Lett.* **2012**, *12*, 161.

[18] R. Y. Tay, M. H. Griep, G. Mallick, S. H. Tsang, R. S. Singh, T. Tumlin, E. H. Teo, S. P. Karna, *Nano Lett.* **2014**, *14*, 839.

[19] L. Wang, B. Wu, J. Chen, H. Liu, P. Hu, Y. Liu, *Adv. Mater.* **2014**, *26*, 1559.

[20] S. Frueh, R. Kellett, C. Mallery, T. Molter, W. S. Willis, C. King'ondu, S. L. Suib, *Inorg. Chem.* **2011**, *50*, 783.

[21] A. Reina, H. Son, L. Jiao, B. Fan, M. S. Dresselhaus, Z. Liu, J. Kong, *J. Phys. Chem. C* **2008**, *112*, 17741.

[22] D. Pacilé, J. C. Meyer, C. O. Girit, A. Zettl, *Appl. Phys. Lett.* **2008**, *92*, 133107.

[23] R. V. Gorbachev, I. Riaz, R. R. Nair, R. Jalil, L. Britnell, B. D. Belle, E. W. Hill, K. S. Novoselov, K. Watanabe, T. Taniguchi, A. K. Geim, P. Blake, *Small* **2011**, *7*, 465.

[24] X. Blase, A. Rubio, S. Louie, M. Cohen, *Phys. Rev. B* **1995**, *51*, 6868.

[25] M. Vanin, J. J. Mortensen, A. K. Kelkkanen, J. M. Garcia-Lastra, K. S. Thygesen, K. W. Jacobsen, *Phys. Rev. B* **2010**, *81*.

[26] S. Joshi, D. Ecija, R. Koitz, M. Iannuzzi, A. P. Seitsonen, J. Hutter, H. Sachdev, S. Vijayaraghavan, F. Bischoff, K. Seufert, J. V. Barth, W. Auwarter, *Nano Lett.* **2012**, *12*, 5821.

[27] P. Sutter, J. Lahiri, P. Albrecht, E. Sutter, *ACS Nano* **2011**, *5*, 7303.

[28] G. Kim, A. R. Jang, H. Y. Jeong, Z. Lee, D. J. Kang, H. S. Shin, *Nano Lett.* **2013**, *13*, 1834.

[29] A. M. van der Zande, P. Y. Huang, D. A. Chenet, T. C. Berkelbach, Y. You, G.-H. Lee, T. F. Heinz, D. R. Reichman, D. A. Muller, J. C. Hone, *Nat. Mater.* **2013**, *12*, 554.

[30] J.-H. Lee, E. K. Lee, W.-J. Joo, Y. Jang, B.-S. Kim, J. Y. Lim, S.-H. Choi, S. J. Ahn, J. R. Ahn, M.-H. Park, C.-W. Yang, B. L. Choi, S.-W. Hwang, D. Whang, *Science* **2014**, *344*, 286.

[31] X. Zhang, Z. Xu, L. Hui, J. Xin, F. Ding, *J. Phys. Chem. Lett.* **2012**, *3*, 2822.

[32] L. Gao, J. R. Guest, N. P. Guisinger, *Nano Lett.* **2010**, *10*, 3512.

[33] Y. Liu, X. Zou, B. I. Yakobson, *ACS Nano* **2012**, *6*, 7053.

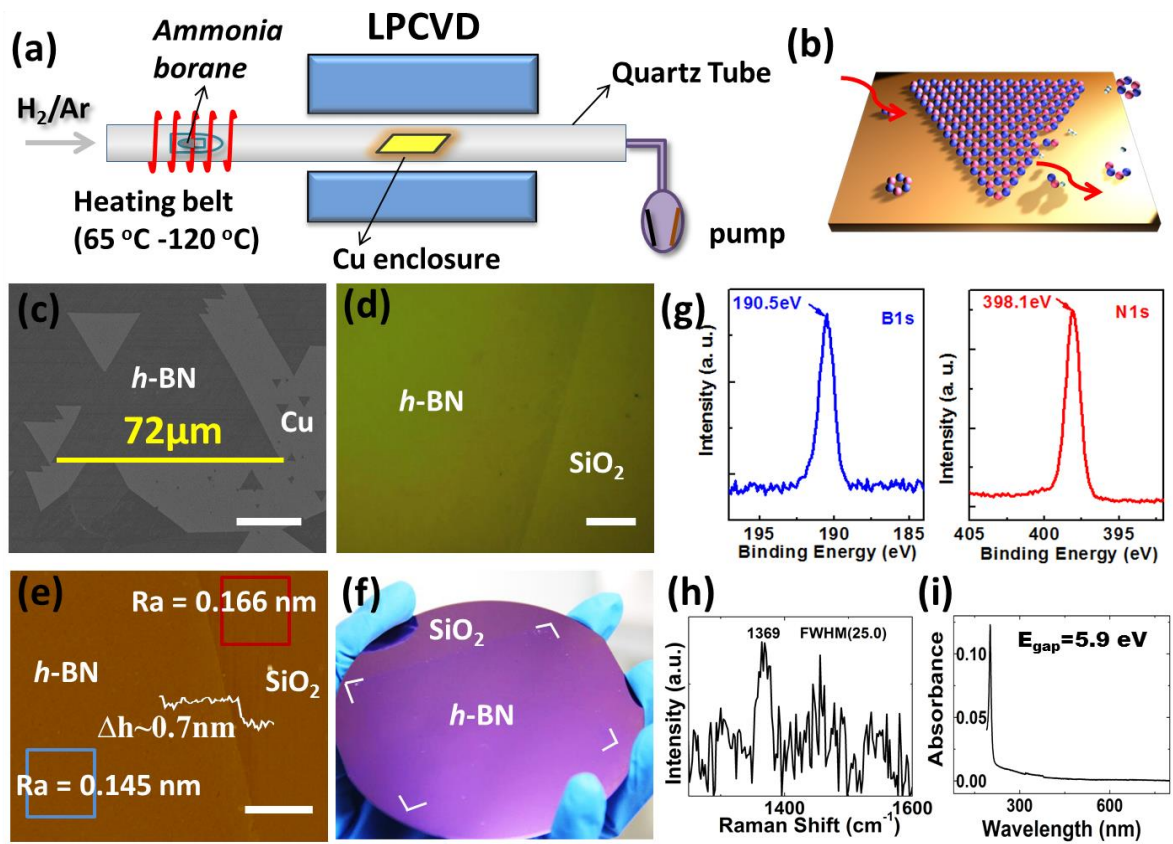

**Figure 1** Synthesis and characterization of *h*-BN films. (**a**) Experimental setup of LPCVD. (**b**) Schematic diagram of the formation of *h*-BN flake on Cu foils. (**c**) SEM image of large domain *h*-BN triangle showing the edge length of ~72 μm. Scale bar: 20 μm. (**d**) Optical microscope image of transferred *h*-BN film on $SiO_2$/Si. Scale bar: 20 μm. (**e**) AFM image of *h*-BN film transferred onto $SiO_2$/Si. The white line profile shows a typical thickness of monolayer *h*-BN on $SiO_2$ (~0.7 nm); the red and blue rectangles indicate the roughness-measuring area at $SiO_2$ and *h*-BN, respectively. Scale bar: 2 μm. (**f**) Wafer-scale monolayer *h*-BN film. (**g**) XPS spectra of B 1s (left) and N 1s (right) with binding energy peaks at 190.5 eV and 398.1 eV, respectively. (**h**) Raman spectrum of *h*-BN on $SiO_2$/Si with a typical peak at 1370 cm$^{-1}$. (**i**) UV-Vis absorption spectrum of *h*-BN film with its obtained band gap value of 5.9 eV.

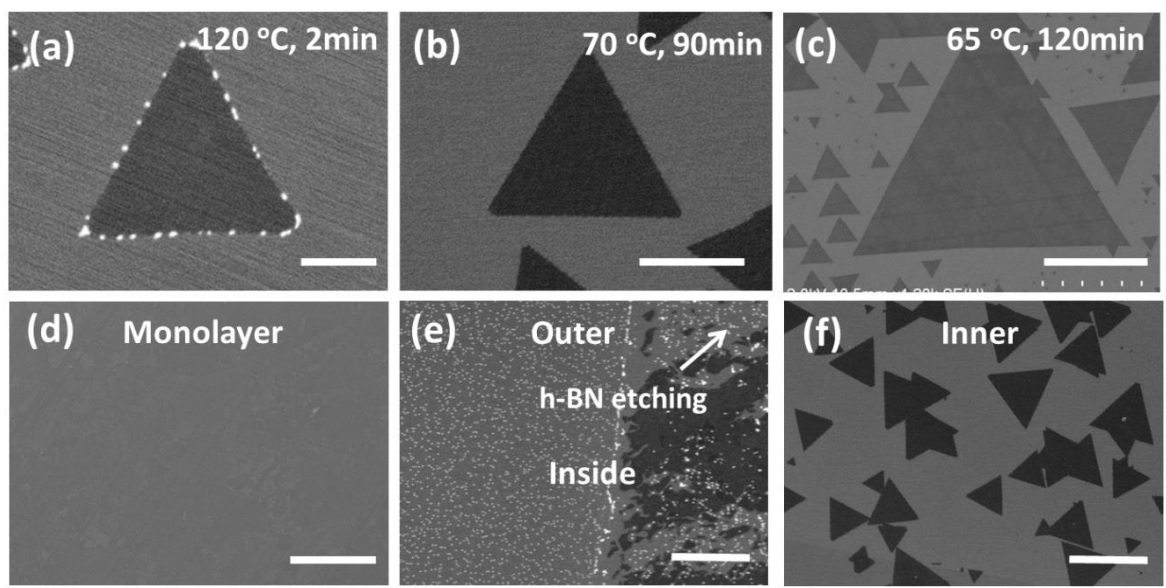

**Figure 2** Effect of precursor evaporation temperature on the CVD growth of *h*-BN triangles with Cu enclosure approach. (**a-c**) SEM images of *h*-BN triangles grown at different precursor heating temperatures and growth times: (**a**) 120 °C, 2 min; (**b**) 70 °C, 90 min; (**c**) 65 °C, 120 min, respectively. (**d**) SEM image of a continuous monolayer *h*-BN obtained as in (c). (**e, f**) SEM micrographs of *h*-BN films obtained as in (**b**) on the outer (**e**) and inner (**f**) surfaces of Cu enclosure, respectively. Scale bar: (**a**) 1μm; (**b**) 5 μm; (**c**) 20 μm; (**d**) 50μm; (**e, f**) 10 μm.

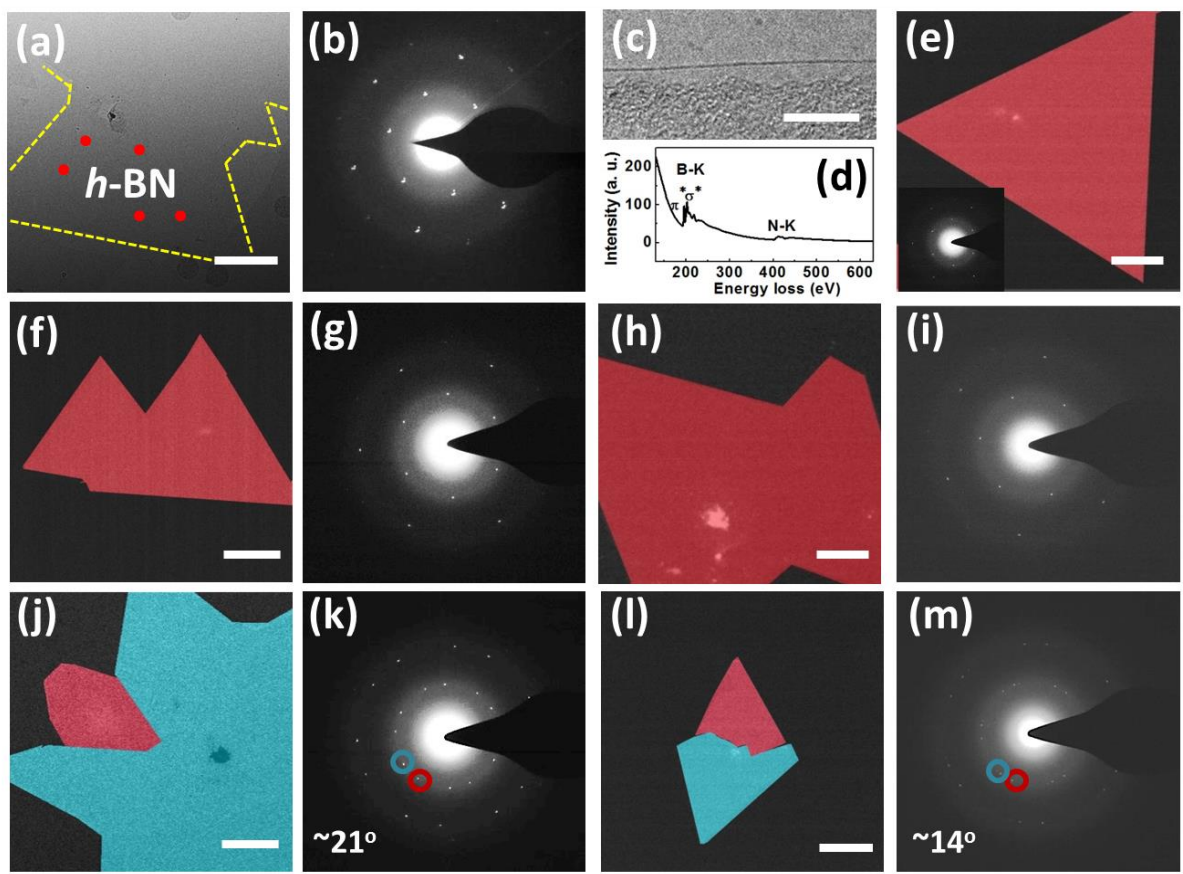

**Figure 3** TEM characterization of *h*-BN film. (**a**) A BF-TEM image of *h*-BN triangle. Scale bar: 10 μm. (**b**) Overlaid SAED pattern randomly taken on *h*-BN triangle shown in (**a**). (**c**) HR-TEM image showing the monolayer nature of *h*-BN film. Scale bar: 10 nm. (**d**) EELS spectrum of the *h*-BN film. (**e**) A false color DF-TEM image of an *h*-BN triangle. Scale bar: 2 μm. (**f**) A false color DF-TEM image of two triangles merged with the same orientation. Scale bar: 2 μm. (**g**) Corresponding SAED pattern for (f). (**h**) A false colored DF-TEM image of the mirror-twin with its corresponding SAED pattern in (**i**). Scale bar: 2 μm. (**j, l**) Other two DF-TEM images of merged *h*-BN flakes with a relative rotation of ~21° and ~14°, respectively. Scale bar: 2 μm. (**k, m**) Corresponding overlaid SAED patterns for (j) and (l), respectively.

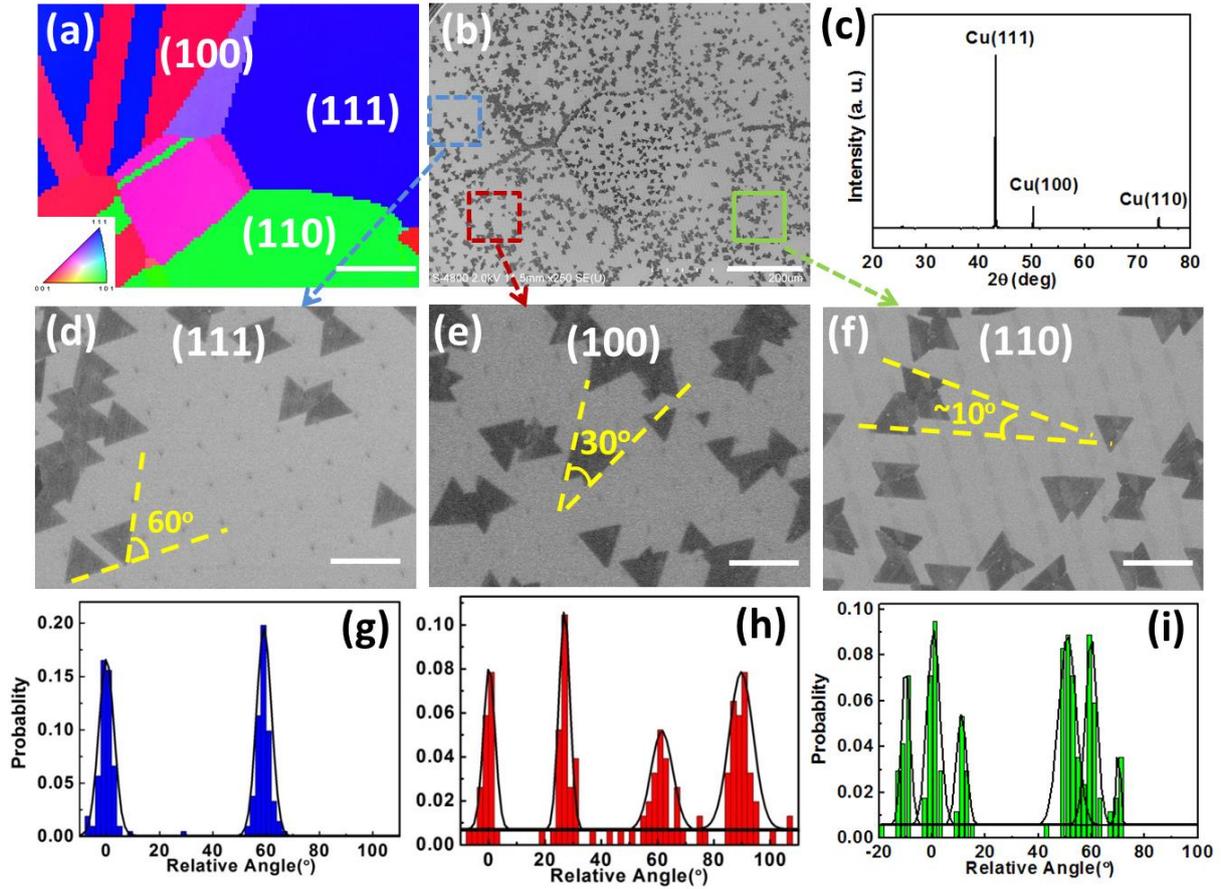

**Figure 4** Orientation dependence of *h*-BN triangle on Cu crystalline facet. (**a**) EBSD mapping of polycrystalline Cu foil. Scale bar: 100 μm. (**b**) Corresponding SEM image of the as-grown *h*-BN on polycrystalline Cu foils. Scale bar: 100 μm. (**c**) X-ray diffraction pattern of Cu foil after growth, consisting of three facets: Cu (111), Cu (100), Cu (110). (**d-f**) Representative SEM images of *h*-BN grown on Cu (111), Cu (100) and Cu (110), respectively. Scale bar: 10 μm. (**g-i**) Statistical distributions of the edge angles of individual triangular *h*-BN domains grown on Cu (111), Cu (100) and Cu (110), respectively.

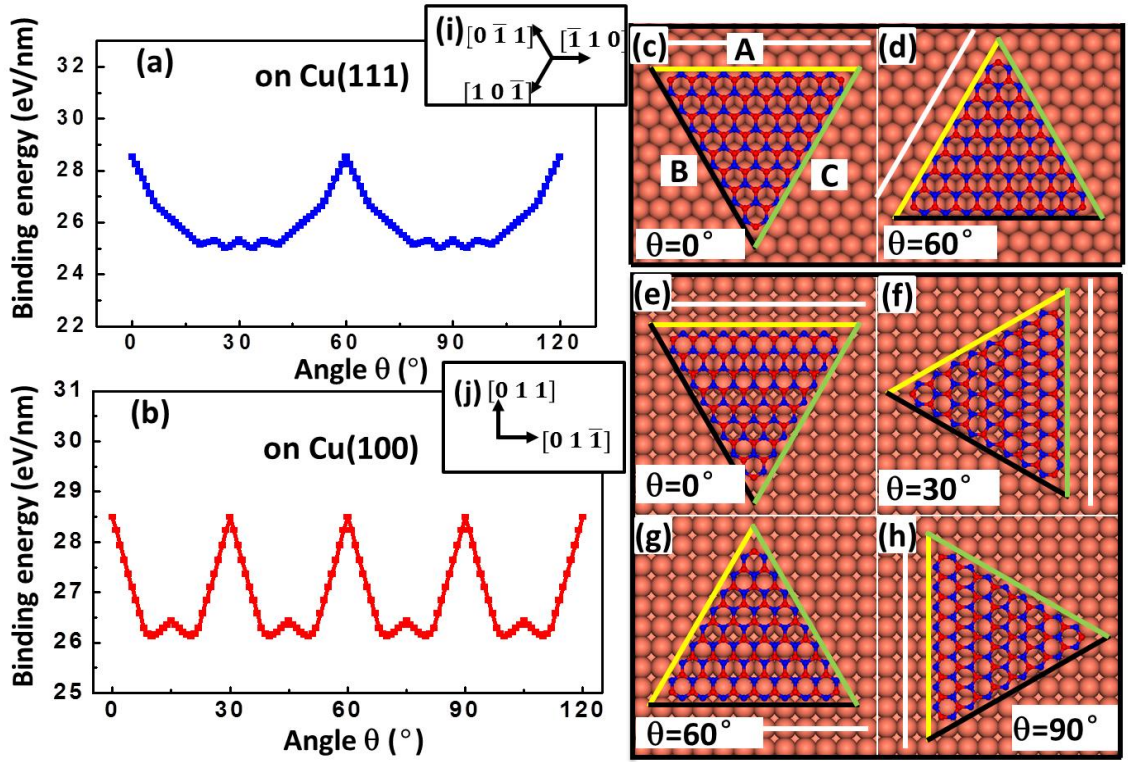

**Figure 5** DFT calculation of the binding energy between *h*-BN flake and Cu substrate. **(a) and (b)** Calculated binding energies between *h*-BN flakes and Cu substrates as a function of angle to the close-packing directions of Cu (111) and Cu (100), respectively. **(c)-(h)** Schematic orientations corresponding to two maximum binding energies of *h*-BN flakes on Cu (111) (**c, d**) and four binding energies peaks on Cu (100) (**e-h**), respectively. The insert images show the three equivalent close-packing directions of Cu (111) (**i**) and two equivalent close-packing directions of Cu (100) (**j**).